\documentclass[aps,prb,twocolumn,superscriptaddress,showpacs,floatfix]{revtex4}

\usepackage{graphicx}

\begin{document}


  \title{Quantum corrections to the conductivity and Hall coefficient of a
two-dimensional electron gas in a dirty AlGaAs/GaAs/AlGaAs quantum
well:\\ from the diffusive to the ballistic regime}



\author{V.~T.~Renard}
\affiliation{GHMFL, MPI-FKF/CNRS, BP-166, F-38042, Grenoble Cedex9, France}
\affiliation{INSA-Toulouse, 31077, Cedex 4, France}
\email{renard@grenoble.cnrs.fr}

\author{I.~V.~Gornyi}
\affiliation{A.~F.~Ioffe Physico-Technical Institute, 194021 St.~Petersburg, Russia}
\affiliation{Institut f\"ur Nanotechnologie, Forschungszentrum Karlsruhe, 76021 Karlsruhe, Germany}

\author{O.~A.~Tkachenko}
\affiliation{Institute of Semiconductor Physics, Novosibirsk 630090, Russia}
\affiliation{GHMFL, MPI-FKF/CNRS, BP-166, F-38042, Grenoble Cedex9, France}

\author{V.~A.~Tkachenko}
\affiliation{Institute of Semiconductor Physics, Novosibirsk 630090, Russia}

\author{Z.~D.~Kvon}
\affiliation{Institute of Semiconductor Physics, Novosibirsk 630090, Russia}
\affiliation{GHMFL, MPI-FKF/CNRS, BP-166, F-38042, Grenoble Cedex9, France}

\author{E.~B.~Olshanetsky}
\affiliation{Institute of Semiconductor Physics, Novosibirsk 630090, Russia}
\affiliation{GHMFL, MPI-FKF/CNRS, BP-166, F-38042, Grenoble Cedex9, France}

\author{A.~I.~Toropov}
\affiliation{Institute of Semiconductor Physics, Novosibirsk 630090, Russia}

\author{J.-C.~Portal}
\affiliation{GHMFL, MPI-FKF/CNRS, BP-166, F-38042, Grenoble Cedex9, France}
\affiliation{INSA-Toulouse, 31077, Cedex 4, France}
\affiliation{Institut Universitaire de France, Toulouse, France}

\date{may 19, 2005}

  \begin{abstract}
We report an experimental study of quantum
conductivity corrections in a low mobility, high density two-dimensional
electron gas in a AlGaAs/GaAs/AlGaAs quantum well in a wide
temperature range (1.5\,K -- 110\,K). This temperature range covers
both the diffusive and the ballistic interaction regimes for our
samples. It has been therefore possible to study the crossover
between these regimes for both the
longitudinal conductivity and the Hall effect.  We perform a
parameter free comparison of our experimental data for the
longitudinal conductivity at zero magnetic field,
the Hall coefficient, and the
magnetoresistivity to the recent theories of interaction-induced
corrections to the transport coefficients. A quantitative
agreement between these theories and our experimental results has been
found.
\end{abstract}


\maketitle

\section{Introduction}
  At low temperatures the conductivity of a degenerated two-dimensional
electron gas (2DEG) is governed by quantum corrections to the
Drude conductivity $\sigma_D$. In general, these corrections have
two principal origins: the weak localization (WL) and the
electron-electron (\mbox{e-e}) interaction \cite{Lee}. Until recently our
understanding of the interaction corrections to the conductivity
of a 2DEG was based on the seemingly unrelated theories developed
for two opposite regimes: the diffusive regime\cite{Altshuler}
$k_BT\tau/\hbar\ll1$, and the ballistic regime\cite{Gold}
$k_BT\tau/\hbar\gg1$. In the diffusive regime the quasi-particle
interaction time $\hbar/k_BT$ is larger than the momentum
relaxation time $\tau$ and two interacting electrons experience
multiple impurity scattering. In the ballistic regime the e-e
interaction is mediated by a single impurity.

Recently, Zala,
Narozhny, and Aleiner (ZNA) have developed a new theory of the
interaction related corrections to the conductivity
\cite{Zala,Zala1} that bridges the gap between the two theories
known previously \cite{Altshuler,Gold}. One of the important
conclusions of the new theory is that the interaction corrections
to the conductivity in both regimes have a common origin: the
coherent scattering of electrons by Friedel oscillations.
This can be also reformulated in terms of returns (diffusive and ballistic)
of an electron to a spatial region which it has already visited.
Conformably to the previous results \cite{Altshuler,Gold}, the new
theory predicts a logarithmic temperature dependence of the
longitudinal conductivity and the Hall coefficient
in the diffusive regime, whereas in the ballistic regime the temperature
dependence of
these parameters becomes linear and $T^{-1}$
respectively. Finally a further step in the generalization of the
interaction theory was realized in Ref.~\onlinecite{Gornyi} and
Ref.~\onlinecite{Sergeev} who considered application of strong
perpendicular magnetic fields for arbitrary type of disorder potential
and influence of electron-phonon
impurity scattering, respectively.

Despite a surge of experimental activity
\cite{Coleridge,Shashkin,Proskuryakov,Kvon,Olshanetsky,Yasin,vitkalov,Pudalov}
following the publication of the theory \cite{Zala,Zala1}, to our
knowledge, no experiment has been reported where the crossover
between the two regimes would have been clearly observed. One of
the reasons is that the temperature at which the crossover is
expected to occur is given by
$k_BT\tau/\hbar\approx0.1$ (see Refs.~\onlinecite{Zala,Zala1,Gornyi}). In the
relatively high-mobility 2D systems that are commonly studied the
crossover temperature is by far too low to be accessed
experimentally ($T<100$\,mK for $\tau>10^{-11}$\,s). Thus, the ZNA
theory has to our knowledge been verified only in the intermediate
and ballistic regimes \cite{Galaktionov}
($k_BT\tau/\hbar=0.1\text{--}10$).

To shift the crossover to higher temperatures one should use low
mobility samples (small $\tau$). At the same time high carrier
densities $N_s$ are necessary in order to maintain high
conductivity and avoid strong localization.
Note that
such samples were already grown and studied
\cite{Kulbachinskii,Minkov,Minkov2} in the diffusive regime,
but the crossover between the ballistic and diffusive regimes
was not considered. In high
density 2D systems the characteristic parameter
$r_s=E_C/E_F\propto1/N_s^{1/2}$, the ratio between Coulomb energy
and Fermi energy is small ($r_s<1$) and hence the effect of e-e
interaction is relatively weak.
In this case the ZNA theory \cite{Zala} predicts insulating like behavior of conductivity
$d \sigma_{xx} / dT > 0$ at all temperatures, whereas the ``screening''
theory \cite{Gold}
predicts metallic like behavior $d \sigma_{xx} / dT < 0$ in the
high-temperature ballistic regime.
Moreover, for such small $r_s$ the Fermi liquid
interaction constant $F_0^\sigma$, the only parameter in the
expressions for the quantum corrections to the conductivity in the
theory \cite{Zala}, can be calculated explicitly.

In this respect low-mobility high-density systems appear to offer
certain advantages for testing the theory \cite{Zala,Zala1}, as
compared to high-mobility low-density systems. Indeed not only
they provide an opportunity for studying an experimentally
accessible temperature crossover between the diffusive and the
ballistic interaction regimes but also the comparison between the
theory and experiment requires no fitting parameters.
Also, in such systems the disorder potential is mostly due to
the short-range impurities which yields almost isotropic
scattering on impurities as assumed in Refs.~\onlinecite{Zala,Zala1}.
As shown in Ref.~\onlinecite{Gornyi}, the nature of disorder becomes crucially
important in the ballistic regime.
Finally, the interaction-induced
longitudinal magnetoresistance  (MR) $\rho_{xx}(B,T)$
in the ballistic regime has been already
studied on systems with long-range \cite{Li} and mixed \cite{Olshanetsky}
disorder, where the theoretical results of Ref.~\onlinecite{Gornyi}
have been confirmed. However, no experimental results for
$\rho_{xx}(B,T)$ have been reported so far for low-mobility systems in
the ballistic regime.

The aim of the present work is to experimentally study the
interaction related corrections to the conductivity, magnetoresistivity,
and the Hall
coefficient in a broad temperature range covering both the
diffusive and ballistic interaction regimes and the crossover
between them. The experimental results obtained in the weak
interaction limit are expected to allow for a parameter free
comparison with the ZNA theory for both $\sigma_{xx}$ and $\rho_{xy}$.
We also compare our results on
the MR for short-range disorder
with the predictions of Ref.~\onlinecite{Gornyi}.

\section{Experimental setup}

  The experimental samples had a 2DEG formed in a narrow ($5$\,nm)
AlGaAs/GaAs/AlGaAs quantum well $\delta$-doped in the middle. Such
doping results in a low mobility and a high carrier density.
Also impurities located in the layer give rise to a large-angle
scattering of electrons. A
detailed description of the structure can be found in
Ref.~\onlinecite{Kvon1}. Two samples from the same wafer have been
studied for which similar results were obtained. Here we present
the data obtained for one of the samples with the following
parameters at $T=1.4$\,K depending on prior illumination: the
electron density $N_s=(2.54-3.41)\times10^{12}$ cm$^{-2}$ and the
mobility $\mu=(380-560)$\,cm$^2$/Vs. The Hall bar shaped samples
were studied between 1.4\,K and 110\,K in magnetic fields up to 15\,T
using a superconducting magnet and a VTI cryostat and also a flow
cryostat ($T>5$\,K) placed in a 20\,T resistive magnet. The data was
acquired via a standard four-terminal lock-in technique with the
current $10$\,nA.

\begin{figure}
\begin{center}
\includegraphics*[angle=-90,width=0.9\columnwidth]{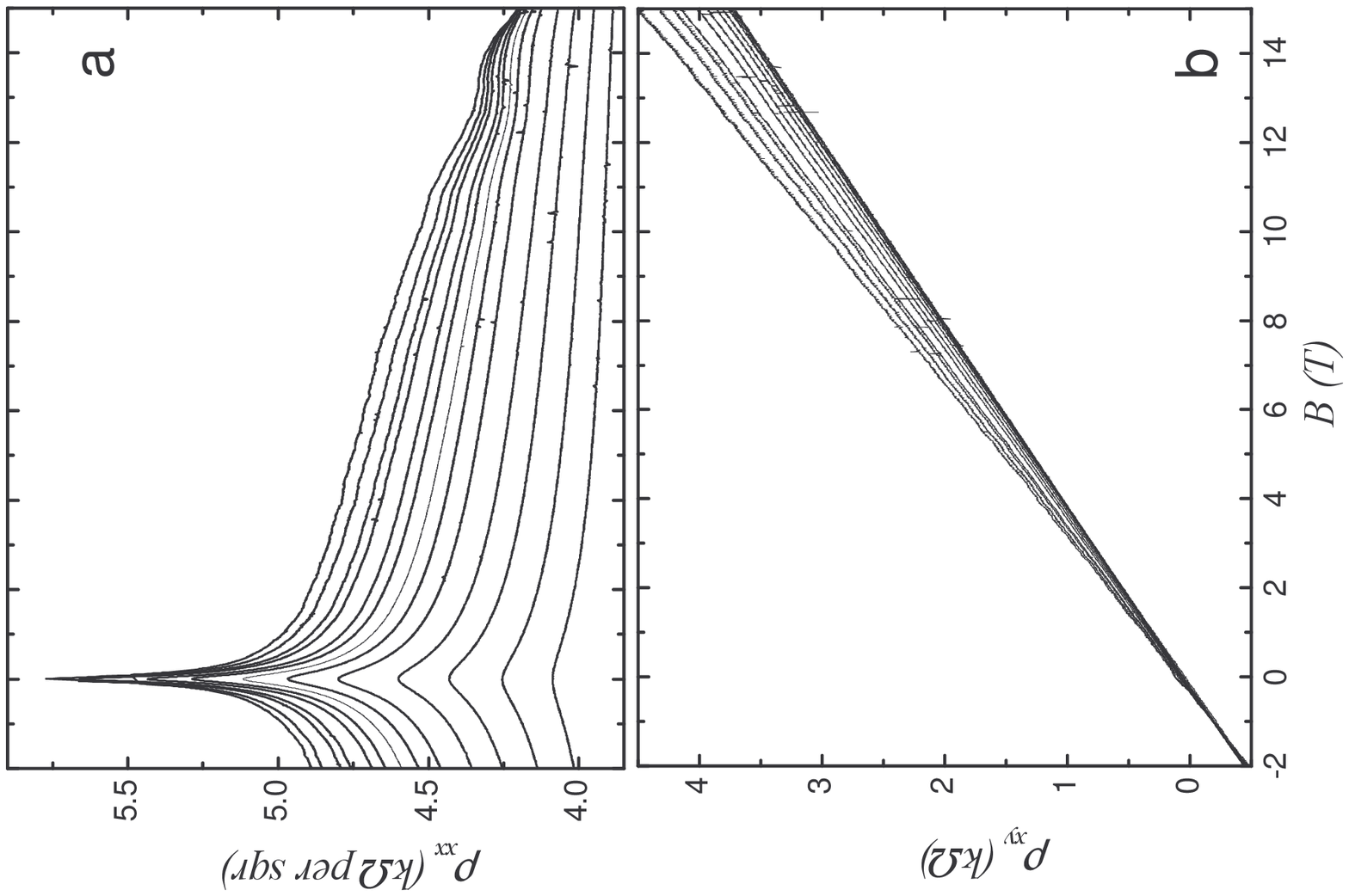}%
\caption{\label{Figure1} a) Longitudinal resistivity of the sample
at $N_s=2.56\times10^{12}$ cm$^{-2}$ for temperature=1.4\,K, 1.9\,K,
3.1\,K, 4\,K, 7.2\,K, 10.25\,K, 15.45\,K, 21.5\,K, 31\,K, 46.2\,K, 62.8\,K,
84.5\,K and 110\,K from top to bottom. b) Hall resistance at the
same temperatures (from top to bottom).}
\end{center}
\end{figure}

  Fig.~\ref{Figure1} shows the longitudinal and Hall resistances of the
sample as a function of magnetic field at temperatures up to
110\,K. As can be seen both are strongly temperature dependent.
Before analyzing the role of the quantum corrections in the
behavior of the transport coefficients shown in
Fig.~\ref{Figure1}, let us estimate the possible contribution from
other unrelated temperature dependent factors.

  First, since the measurements were performed up to relatively high
temperatures, the question of the role of phonon scattering
becomes important. In this connection we believe that the
following argument can be used. It is well known that in
ultra-clean GaAs samples sufficiently high values of mobility are
reported even at liquid nitrogen temperatures (see, for example
Ref.~\onlinecite{Pfeiffer,Lin}, where $\mu=4\times10^5$ cm$^2$/Vs at
$T=77$\,K). At these temperature the phonon scattering is the
dominant scattering mechanism in these samples and yet the
mobilities are still a thousand times larger than in our sample.
In our experiment, the pure electron-phonon contribution to the
conductivity is thus negligible compared to impurity
scattering.

Recently a theory of the interplay between electron-phonon and impurity
scattering was developed\cite{Sergeev}. It was argued that these
interference effects might play a significant role at intermediate
temperatures. However, we have evaluated the phonon contribution to be a
few percent of the Drude conductivity at 100\,K. Also our estimates show that
this contribution is still smaller than the e-e interaction one.
For these reasons
the effect of phonons can be neglected in the entire experimental
temperature range in these samples.

Now, as can be seen in Fig.~\ref{Figure1}, the Hall coefficient
varies with $T$ at low
temperatures but remains practically constant for $T>20$\,K. One
might argue that the behavior at low temperatures could be due to
a variation of the electron density with temperature. However, we
believe that this is not the case. Indeed, from the measurements
carried out up to 20\,T where the Shubnikov - de Haas (SdH) oscillations
are better resolved, we find that the density remains constant at
$T<30$\,K. Also we find that the density given by the SdH
oscillations is the same as we get from the slope of the Hall
resistance at $T>20$\,K where it is $T$-independent. We conclude
therefore that the electron density remains constant in the entire
experimental temperature range and all the data presented in
Fig.~\ref{Figure1} corresponds to $N_s=2.56\times10^{12}$
cm$^{-2}$.

  Having excluded the phonon scattering and the density variation as
possible causes of the behavior shown in Fig.~\ref{Figure1} we
associate the observed temperature dependencies with the quantum
corrections to the transport coefficients. Our data will be
analyzed in the framework of the recent theories \cite{Zala,Zala1}
valid for a degenerated 2DEG ($k_BT\ll E_F$). According to
Ref.~\onlinecite{Kvon1} only one subband is occupied in our
quantum wells at $N_s=2.56\times10^{12}$ cm$^{-2}$. Also
$E_F\approx1000$\,K and so the theory \cite{Zala,Zala1} should
apply under our experimental conditions.

\section{Quantum corrections to the conductivity tensor:
background}
\label{background}

The longitudinal conductivity is a sum of three components:
the Drude conductivity, the WL contribution and the e-e
interaction correction. For the correct evaluation of the
interaction related correction at $B=0$~T, the knowledge of the
first two contributions to the conductivity is essential.
Unfortunately, in our case there is no direct means of knowing the
value of the Drude conductivity $\sigma^D_0$ because of a
considerable (up to $20\%$) variation of the zero field
conductivity with temperature. On the other hand, to single out
the e-e interaction correction we have to eliminate the weak localization
contribution, which might be stronger than the interaction correction at $B=0$.

The WL correction to the conductivity
at low temperatures and magnetic fields
is described by a well-known expression~\cite{Hikami}
involving digamma functions.
However, at high enough temperatures and/or magnetic fields
(when the contribution of non-diffusive paths
becomes more and more pronounced) the WL
correction is given by a rather complicated analytical
expression~\cite{mcphail}.
Nevertheless, there exists a method (see the next Section)
that can be used for the evaluation of all the three contributions
to the conductivity at zero magnetic field basing on the knowledge
of the high-B behavior of the magnetoconductivity. This method has the
advantage that one can forgo the usual procedure of fitting the
low field data with the theoretical expressions for the WL
magnetoresistance\cite{Hikami,mcphail}, thus eliminating a possible
source of error at this stage.

A general formula for the conductivity tensor in a magnetic
field can be derived using the quantum kinetic
equation of Ref.~\onlinecite{Zala}.
The longitudinal and the Hall conductivities can be written for $k_BT\ll E_F$
in the following form~\cite{Gornyi1}
\begin{eqnarray}
\sigma_{xx}(T,B)&=&\frac{\sigma_D(T)}{1+\omega_c^2\tau^2(T)}+\Delta\sigma^\mathrm{Diff}_{ee}(T)\nonumber \\
&&\hskip3cm +\
\Delta\sigma_{xx}^{\mathrm{WL}}(T,B),\nonumber \\
\label{Eq1}
\\
\sigma_{xy}(T,B)&=& \frac{\omega_c\tau(T)\sigma_D(T)}{1+\omega_c^2\tau^2(T)}
+ \omega_c\tau(T)\Delta\sigma^\mathrm{H}_{ee}(T)\nonumber \\
&&\hskip3cm +\ \Delta\sigma_{xy}^{\mathrm{WL}}(T,B),\nonumber
\\
\label{Eq1-xy}
\end{eqnarray}
Generally, the zero-$B$ Drude conductivity $\sigma_D(T)$ depends
on $T$ due to the interaction-induced renormalization of both the
transport scattering time $\tau(T)$ and the Fermi velocity
$v_F(T)$. Strictly speaking, the cyclotron frequency $\omega_c$
also depends on $T$ via the renormalization of the effective mass
$m^*(T)$; however, it appears in (\ref{Eq1}) and (\ref{Eq1-xy})
only in combination $\omega_c\tau(T)$ so that one can absorb its
renormalization into the $T$-dependence of the effective
scattering time. While the first terms in (\ref{Eq1}) and
(\ref{Eq1-xy}) have the structure of the classical Drude
conductivity in a finite $B$, the terms
$\Delta\sigma^\mathrm{Diff}_{ee}(T)$ and
$\omega_c\tau(T)\Delta\sigma^\mathrm{H}_{ee}(T)$ appear as quantum
corrections to the Drude terms.

The expressions (\ref{Eq1}) and (\ref{Eq1-xy}) are justified under the condition
\begin{equation}
\omega_c\ll \pi/\tau+2\pi^2 k_B T/\hbar
\label{sdho}
\end{equation}
which allows one to neglect SdH oscillations in the
present case of short-range impurity potential. The same condition
governs the strength of the influence of magnetic field on the
collision integral in the kinetic equation~\cite{Zala} and allows
one to neglect cyclotron returns to the same impurity. Under this
condition, the bending of relevant electron trajectories by the
magnetic field is weak. It is taken into account by a proper
definition of the the $B$-independent quantities $\sigma_D(T),$
$\tau(T),$ $\Delta\sigma^\mathrm{Diff}_{ee}(T),$ and
$\Delta\sigma^\mathrm{H}_{ee}(T).$ This makes it
possible~\cite{Gornyi1} to extract the interaction-induced
corrections to the conductivity at $B=0$ from the
magnetoconductivity obtained in relatively strong magnetic fields,
see Section~\ref{method}. The  condition (\ref{sdho}) is fulfilled
in the whole range of relevant $T$ and $B$ we address in this
work.

The term $\Delta\sigma^\mathrm{Diff}_{ee}$ in Eq.~(\ref{Eq1})
corresponds to the ``diffusive'' contribution of e-e interactions,
which is due to the coherent processes involving multiple impurity
scattering events. In the diffusive regime,
$\Delta\sigma^\mathrm{Diff}_{ee}$ diverges
logarithmically~\cite{Altshuler} with decreasing $T$,
$\Delta\sigma^\mathrm{Diff}_{ee}(T)\propto \ln(k_B T\tau/\hbar)$
[we note in passing that, in contrast to
Eqs.~(\ref{ZNA-delta-sigma-C}) and (\ref{ZNA-delta-sigma-T})
below, this logarithmic contribution is cut off by $\hbar/\tau$
rather than by $E_F$]. At high temperatures the contribution of
diffusive paths is expected to vanish, since the probability of
``diffusive'' returns involving more than one impurity-scattering
is suppressed in the ballistic regime (each additional impurity
scattering yields an extra factor $\hbar/k_B T\tau$). In effect,
the term $\Delta\sigma^\mathrm{Diff}_{ee}$ in (\ref{Eq1}) also
takes into account the influence of the magnetic field on the
return probability determining the correction to the $T$-dependent
part of the effective transport time, see discussion in
Ref.~\onlinecite{Gornyi}. This contribution to
$\Delta\sigma^\mathrm{Diff}_{ee}$ dominates in the ballistic
regime. As for the term $\Delta\sigma^\mathrm{H}_{ee}$ in
Eq.~(\ref{Eq1-xy}), the contribution of diffusive paths to it is
exactly zero~\cite{Altshuler}, so that this term is completely
determined by the influence of the magnetic field on the collision
integral. Therefore, in the diffusive regime
$\Delta\sigma^\mathrm{H}_{ee}$ has no logarithmic
divergency~\cite{Altshuler}, unlike
$\Delta\sigma^\mathrm{Diff}_{ee}$.

Taking into account the
e-e interaction effects related  to the scattering on a single impurity results in
the $T$-dependent renormalization \cite{Gold,Zala,Gornyi1} of $\tau(T)$.
The $T$-dependence of $\sigma_D(T)$ in the ballistic limit
is dominated by the $T$-dependence of $\tau(T)$ since the interaction-induced
correction to the Fermi velocity yields a weaker $T$-dependence.
Thus, in the ballistic limit the linear-in-$T$ interaction
correction to the zero-$B$ conductivity~\cite{Zala} enters Eq.~(\ref{Eq1}) only via the
renormalized transport scattering time $\tau(T)$ in the first term
(both in the numerator and the denominator).

The terms $\Delta\sigma^{\mathrm{WL}}_{xx}$ and $\Delta\sigma^{\mathrm{WL}}_{xy}$ in
Eqs.~(\ref{Eq1}) and (\ref{Eq1-xy}) are the WL corrections to the
longitudinal and Hall conductivities, respectively. Actually, the
WL corrections arise solely from the renormalization of the
transport scattering time~\cite{DKG} and hence they can be
completely absorbed into the first ``classical'' terms in
(\ref{Eq1}) and (\ref{Eq1-xy}) via the $B$-dependent correction to
$\tau(T)$.

A general method for the analysis of the magnetotransport data is
based on Eqs.~(\ref{Eq1}) and (\ref{Eq1-xy}).  For a given
temperature one can treat the $B$-independent quantities
$\sigma_D(T),$ $\tau(T),$ $\Delta\sigma^\mathrm{Diff}_{ee}(T),$
and $\Delta\sigma^\mathrm{H}_{ee}(T)$ as four fitting parameters
to fit the two experimental curves: $\sigma_{xx}(T,B)$ vs $B$ and
$\sigma_{xy}(T,B)$ vs $B$. Under the assumption that the WL
corrections are suppressed it follows from Eqs.~(\ref{Eq1}) and
(\ref{Eq1-xy}) that
$$
\sigma_{xx}(T,B)={\sigma_{xy}(T,B)\over \omega_c\tau(T)}
+\Delta\sigma^\mathrm{Diff}_{ee}(T)-\Delta\sigma^\mathrm{H}_{ee}(T).
$$
This equation allows one to find the values of $\tau(T)$
from the slope of $\sigma_{xx}(T,B)$ vs $\sigma_{xy}(T,B)/\omega_c$
dependence. Then $\sigma_D(T)$ and
$\Delta\sigma^\mathrm{Diff}_{ee}(T)$ can be found from the
$\sigma_{xx}(T,B)$ vs $[1+\omega_c^2\tau^2(T)]^{-1}$ plot. A
detailed analysis of the magnetoconductivity based on this
procedure will be published elsewhere~\cite{unpub}. In this paper
we will concentrate on the zero-$B$ interaction-induced correction
to $\sigma_{xx}.$ For this purpose a simpler fitting procedure
described in Section~\ref{method} is sufficient.

\section{Longitudinal conductivity at $B=0$: Experimental method}
\label{method}

Let us describe how the experimental quantum corrections
were extracted from the row data and then turn to the analysis of
the obtained corrections. The main idea of our method is to use the
MR and Hall data obtained in a relatively {\it strong magnetic field},
where the weak localization is suppressed,
to find the value of interaction-induced corrections
in the limit of {\it zero magnetic field}.

With the magnetic field increasing, the MR in Fig.~\ref{Figure1}a goes
through two distinct types of behavior. An abrupt drop of
resistance at low fields and then a much weaker magnetic field
dependence at higher $B$.
As is well known the weak localization
is suppressed at magnetic fields larger than
$B_{tr}=\hbar/(2el^2)$, where $l$ is the mean free path. In our
samples $B_{tr}\approx1.5$\,T that roughly coincides with the field
at which the crossover from the one type of MR to the other takes
place. We conclude therefore that the strong MR observed at low
fields can be associated with the WL suppression in our samples
and that the MR observed at higher fields must be attributed
entirely to the e-e interaction effects \cite{Altshuler}.

\begin{figure}
\begin{center}
\includegraphics*[angle=-90,width=0.9\columnwidth]{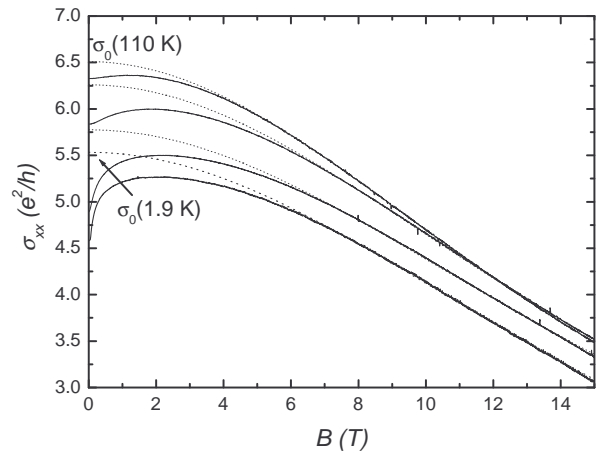}%
\caption{\label{Figure2} Experimental longitudinal conductivity at
$T=1.9$\,K, 10.25\,K, 62.8\,K and 110\,K from bottom to top (solid
line) and the fit to Eq.~(\ref{Eq1}) according to the description
in the text. The result is then extrapolated to $B=0$\,T (dotted
line) for the same temperatures.}
\end{center}
\end{figure}

As a first step of our procedure, the experimental values of the
longitudinal and Hall conductivities are obtained by inverting the
resistivity tensor using the data shown in Fig.~\ref{Figure1}. The
result for the longitudinal conductivity is shown in
Fig.~\ref{Figure2}. The weak localization correction dominates at
low fields but is suppressed at $B>B_{tr}$. Therefore, at $B\gg
B_{tr}$ the shape of the $\sigma_{xx}$ vs $B$ dependence should be
determined by the first term in Eq.~(\ref{Eq1}). The term
$\Delta\sigma^\mathrm{Diff}_{ee}$, which is $B$-independent,
should only result in a vertical shift of this contribution.
At low temperatures we experimentally find that with the WL
suppressed at higher magnetic fields the MC corresponding to
different temperatures forms parallel vertically shifted traces
(see Fig.~\ref{Figure2}) whose shape is given by the first term in
Eq.~(\ref{Eq1}) with a $T$-independent $\tau$. At temperatures
larger than $30$\,K the shape of the curves begins to deviate
slightly from that of the low temperature traces. This change is
attributed to the renormalization of the scattering time by e-e
interactions in the ballistic limit \cite{Zala}.

To interpolate between all the relevant
regimes (diffusive vs ballistic, classically weak $B$ vs strong $B$)
we use a simplified version of Eq.~(\ref{Eq1}).
Within this procedure, we attribute the $T$-dependence of $\sigma_D(T)$
solely to the $T$-dependence of $\tau(T)$, using
\begin{equation}
\sigma_D(T)=e^2 n\tau(T)/m^*.
\label{sigmaD}
\end{equation}
This amounts to treating all the interaction-induced corrections
to the collision integral related to
$\sigma_D(T)$ as the renormalization of the effective
transport scattering time. Further, we assume that the
$T$-dependence of the product $\omega_c\tau$ in the classical terms in
(\ref{Eq1}) and (\ref{Eq1-xy})
is the same as the $T$-dependence of $\sigma_D(T)$.
This approximation [used earlier together with Eq.~(\ref{sigma_xx_no_WL})
in Refs.~\onlinecite{Minkov2,Minkov3}]
yields the
proper asymptotics of the conductivity correction, that are
governed by $\Delta\sigma^\mathrm{Diff}_{ee}$ and $\tau(T)$ in the
diffusive and the ballistic regimes, respectively.

\begin{figure}
\begin{center}
\includegraphics*[angle=-90,width=0.9\columnwidth]{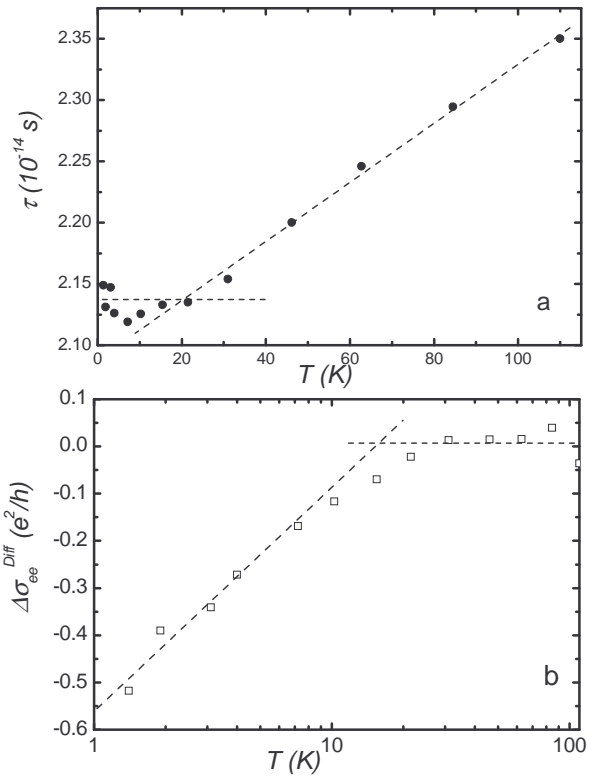}%
\caption{\label{Figure3} Obtained values of the scattering time
(a) and of the term $\Delta\sigma^\mathrm{Diff}_{ee}$ (b). A clear
change in their behavior is observed from constant (logarithmic)
to linear (constant) at $T\sim20$\,K. The lines are a guide for the
eyes.}
\end{center}
\end{figure}

It is possible to determine the scattering time by fitting the
curves for $B>6$\,T using Eq.~(\ref{Eq1}) at a given value of $T$
with $\tau(T)$ and $\Delta\sigma^\mathrm{Diff}_{ee}(T)$ as fitting
parameters (see Fig.~\ref{Figure2}). This was done for all the
temperatures and the results for both
$\Delta\sigma^\mathrm{Diff}_{ee}$ and $\tau(T)$ are presented in
Fig.~\ref{Figure3}.
The momentum relaxation time is observed to increase linearly with
temperature at $T>20$\,K. This linear behavior is expected in the
ballistic limit\cite{Gold,Zala}. As for the term
$\Delta\sigma^\mathrm{Diff}_{ee}$, it is observed to decrease in
amplitude with temperature increasing and to vanish at $T>20$\,K.
It is important to stress that a significant change in the
behavior of these two parameters occurs at $T=20$\,K.

Once fitted for $B>6$\,T, the term
\begin{equation}
{\tilde\sigma}_{xx}(T,B)= \frac{e^2
n}{m^*}\frac{\tau(T)}{1+\omega^2_c\tau^2(T)}+\Delta\sigma^\mathrm{Diff}_{ee}(T)
\label{sigma_xx_no_WL}
\end{equation}
was then extrapolated for each of the curves down to $B=0$ (see
Fig.~\ref{Figure2}). We believe that the value
\begin{equation}\sigma_0(T)={\tilde\sigma}_{xx}(T,B=0)=
\sigma_D(T)+\Delta\sigma^\mathrm{Diff}_{ee}(T) \label{sumdelta}
\end{equation}
obtained at $B=0$ is free of the $T$-dependent WL contribution~\cite{foot-WL}.

Finally, the
temperature independent term $\sigma^D_0$
was subtracted from $\sigma_0$ at
all temperatures. This was made to obtain the value of the e-e
interaction correction to the conductivity
\begin{equation}
\Delta\sigma_{xx}^{ee}(T,B=0)=\sigma_0(T)-\sigma^D_0
\label{difference}
\end{equation}
which is presented in figure \ref{Figure4}. The value
$\sigma^D_0=(6.3\pm 0.1) \times e^2/h$ was found from the analysis of the
MR data as the value of the conductivity at the point
$\omega_c\tau=1,$ where the MR curves corresponding to the
diffusive range of $T$ intersect, see  Section~\ref{MR} and
Eq.~(\ref{Drude-rho}) there.

\section{Longitudinal conductivity at $B=0$: Experiment vs Theory}

According to Ref.~\onlinecite{Zala}, the e-e interaction correction
to the conductivity is given by the following expressions:
\begin{equation}
\label{Eq2} \Delta\sigma_{xx}^{ee}=\delta\sigma_C+3\delta\sigma_T,
\end{equation}
where
\begin{eqnarray}
\delta\sigma_C&=&\frac{e^2}{\pi\hbar}\frac{k_{B}T\tau}{\hbar}\left[1-\frac{3}{8}
f(k_{B}T\tau/\hbar)\right]\nonumber \\
&-& \frac{e^2}{2\pi^2\hbar}\ln\left[\frac{E_F}{k_{B}T}\right]
\label{ZNA-delta-sigma-C}
\end{eqnarray}
is the charge channel correction and
\begin{eqnarray}
\delta\sigma_T&=&\frac{F_{0}^{\sigma}}{\left[1+F_{0}^{\sigma}\right]}
\frac{e^2}{\pi\hbar}\frac{k_{B}T\tau}{\hbar}
\left[1-\frac{3}{8}t(k_{B}T\tau/\hbar;F_{0}^{\sigma})\right]
\nonumber \\[0.2cm]
&-&\left[1-\frac{\ln(1+F_{0}^{\sigma})}{F_{0}^{\sigma}}\right]
\frac{e^2}{2\pi^2\hbar}\ln\left[\frac{E_F}{k_{B}T}\right]
\label{ZNA-delta-sigma-T}
\end{eqnarray}
is the correction in the triplet channel. The detailed expression of
$f(x)$ and $t(x;F_0^\sigma)$ can be found in Ref.~\onlinecite{Zala}.

In these expressions the linear-in-$T$ term is due to the
renormalization of $\tau(T)$ by Friedel oscillation. This
contribution comes from $\sigma_D(T)$ in Eq.~(\ref{sumdelta}) and
dominates in the ballistic limit $k_BT\tau/\hbar\gg 1$. In the
diffusive limit, the conductivity correction is determined by the
logarithmic terms, which can be roughly split in two parts as
follows: $\ln(E_F/k_B T)=\ln(\hbar/k_BT\tau)+\ln(E_F\tau/\hbar).$
Here the first (singular) term comes from
$\Delta\sigma^\mathrm{Diff}_{ee}$ in Eqs.~(\ref{Eq1}) and
(\ref{sumdelta}). The second (constant) term is the contribution
of $\sigma_D(T)$. In the ballistic regime
$\Delta\sigma^\mathrm{Diff}_{ee}$ gets suppressed, so that the
whole logarithmic term $\ln(E_F/k_B T)$ comes from $\sigma_D(T)$.

It is
worth mentioning that for small $r_s$ the interaction constant
$F_0^\sigma$ as function of $r_s$ can be calculated explicitly.
As suggested by ZNA~\cite{Zala} we used:
\begin{equation}
F_0^\sigma\rightarrow-\frac{1}{2}\frac{r_s}{r_s+\sqrt{2}}=-0.1
\label{F-ball}
\end{equation}
in the first line of $\delta\sigma_T$ (this form reflects the backscattering
character of e-e interaction related to Friedel oscillations) and
\begin{equation}
F_0^\sigma\rightarrow-\frac{1}{2\pi}\frac{r_s}{\sqrt{2-r_s^2}}
\ln\left(\frac{\sqrt{2}+\sqrt{2-r_s^2}}{\sqrt{2}-\sqrt{2-r_s^2}}\right)=-0.17
\label{F-diff}
\end{equation}
in the second line so that no additional fitting parameter has
been introduced. The calculations were done for $r_s=0.35$
corresponding to the electron density in our
sample.

\begin{figure}
\includegraphics*[angle=-90,width=\columnwidth]{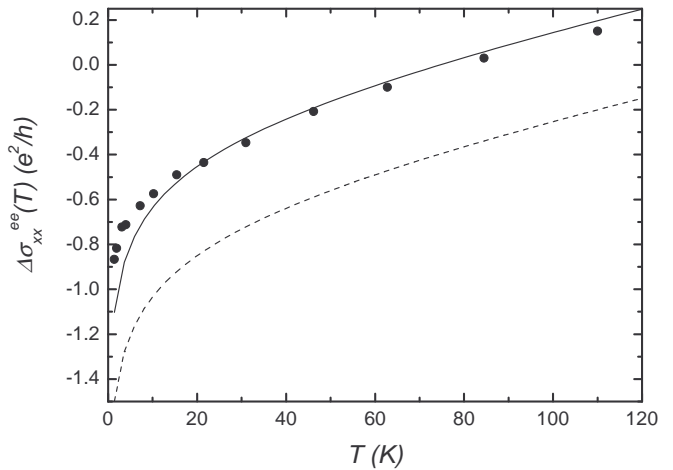}%
\caption{\label{Figure4} Experimental temperature dependence of
the e-e correction to conductivity (black dots). The dashed line
corresponds to the first evaluation of the model of
Ref.~\onlinecite{Zala}. The solid line corresponds to the theory
taking into account the temperature independent contribution (see
the text).}
\end{figure}

In Fig.~\ref{Figure4} we plot the theoretical curve (dashed
line) calculated for our system parameters using Eqs.~(\ref{Eq2}),(\ref{ZNA-delta-sigma-C}),
and (\ref{ZNA-delta-sigma-T}),
as well as the experimental data points. As can be seen,
there is a systematic
shift of the experimental points with respect to the theoretical
curve. This shift can be explained by the fact that ZNA theory describes
only the temperature dependence of the conductivity but not the magnitude
of the total interaction-induced contribution.

Firstly, in addition
to the correction $\Delta\sigma_{xx}^{ee}$ given by
Eqs.~(\ref{Eq2}), (\ref{ZNA-delta-sigma-C}), and (\ref{ZNA-delta-sigma-T}),
there is a large $T$-independent interaction-induced contribution to
conductivity which is due to the $T$-independent part of the
renormalization (screening) of impurities by Friedel oscillations
[see Eq.~(3.33) of Ref.~\onlinecite{Zala}].
For $r_s\agt 1,$ this contribution is of the same order in magnitude
as the value of the Drude conductivity of a noninteracting electron gas,
while for $r_s\ll 1$ it contains an additional factor $\sim r_s$.
However, in the presence of interactions this contribution cannot be
experimentally separated from the noninteracting part of the Drude conductivity.
Therefore, the value of $\sigma^D_0$ used in Eq.~(\ref{difference})
already takes into account this screening-induced term,
so that the observed shift cannot be explained in
this way.

Secondly, the logarithmic terms in (\ref{ZNA-delta-sigma-C})
and (\ref{ZNA-delta-sigma-T}) yield a $T$-independent
contribution which depends on the ultraviolet cutoff.
It is worth noting that $E_F$ appears
in Eqs.~(\ref{ZNA-delta-sigma-C}) and (\ref{ZNA-delta-sigma-T}) only
due to the contribution of $\sigma_D(T)$
(this fact becomes important in a finite magnetic field).
It follows that, similarly to the linear-in-$E_F$ term discussed above,
the $T$-independent term
$\sim \ln(E_F\tau/\hbar)$ is also already absorbed in $\sigma^D_0$ when the latter
is found from the analysis of the MR data.
Thus it is not surprising to observe a vertical shift
between the predictions of Ref.~\onlinecite{Zala} written in the form of
Eqs.~(\ref{ZNA-delta-sigma-C}) and (\ref{ZNA-delta-sigma-T}) with $\ln(E_F/k_B T)$
and the experimental data obtained using a specific choice of the value of the
Drude conductivity.

In Fig.~\ref{Figure4} we shifted the theoretical curve given by
Eqs.~(\ref{ZNA-delta-sigma-C}) and (\ref{ZNA-delta-sigma-T})
upward by replacing $E_F$ by a quantity of order of $\hbar/\tau$
in the logarithmic terms (solid line).


A reasonably good quantitative agreement between the model of
Ref.~\onlinecite{Zala} and the data is found for the entire
temperature range.
Note that contrary to the previous works
\cite{Coleridge,Shashkin,Proskuryakov,Kvon,Olshanetsky,Yasin} we
have used no fitting parameter. Moreover we find that using the
interaction constant $F_0^\sigma$ as a fitting parameter does not
result in a better agreement between theory and experiment.

Let us now return to the analysis of Fig.~\ref{Figure3} which we
believe to reveal important information. Indeed, the total
correction to the conductivity is the sum of
$\Delta\sigma^\mathrm{Diff}_{ee}(T)$ and a ballistic contribution
which is proportional to $\tau(T)$. As can be seen the logarithmic
diffusive part vanishes at $T>20$\,K when the ballistic part starts
to vary linearly with temperature. Therefore we believe that
Fig.~\ref{Figure3} establishes a crossover from the diffusive to
the ballistic limit in the behavior of the interaction-induced
correction to the zero-$B$ conductivity. This change of behavior
is observed at $T\sim 20$\,K. This is in a qualitative agreement with
the ZNA theory, predicting the crossover to occur at
$k_BT\tau/\hbar\sim0.1$ which corresponds in our case to
$T\sim30$\,K.

Finally, not only Fig.~\ref{Figure3} shows that the scattering
time effectively varies linearly with temperature at high
temperature \cite{Gold,Zala} but it also shows that the sign of
the variation is positive (i.e. insulating like). It is due to the
fact that at small $r_s$ the exchange (singlet) contribution is more important
than Hartree (triplet) contribution \cite{Zala}. While predicted by ZNA theory
at low interaction this behavior is not allowed by the screening
theory\cite{Gold} which does not take into account the exchange
part in the calculation of the corrections.

\section{Hall effect}

  We now turn to the analysis of the Hall data presented in
Figure.~\ref{Figure1}b. According to Ref.~\onlinecite{Zala1} the
Hall resistivity may be written as:
\begin{equation}
\rho_{xy}=\rho_H^D+\delta\rho_{xy}^C+\delta\rho_{xy}^T \label{Eq3}
\end{equation}
where $\rho_H^D$ is the classical Hall resistivity and
$\delta\rho_{xy}^C,\delta\rho_{xy}^T$ are the corrections in the
charge and triplet channel. These corrections are given as
follows:
\begin{equation}
\frac{\delta\rho_{xy}^C}{\rho_H^D}=
\frac{1}{\pi E_F\tau}\ln\left(1+\lambda\frac{\hbar}{k_BT\tau}\right)
\label{Eq4}
\end{equation}
\begin{equation}
\frac{\delta\rho_{xy}^T}{\rho_H^D}=
\frac{3 h(k_BT\tau/\hbar;F_0^\sigma)}{\pi E_F\tau}\ln\left(1+\lambda\frac{\hbar}{k_BT\tau}\right)
\label{Eq4t}
\end{equation}
The detailed expression for $h(x;F_0^\sigma)$ can be found in
Ref.~\onlinecite{Zala1}, $\lambda=\frac{11\pi}{192}$ and the value
of $\rho_H^D$ is obtained from the high temperature curves for
which
$\delta\rho_{xy}\rightarrow0$.
Therefore according to the theory of the e-e interaction \cite{Zala1} one should
observe a logarithmic temperature dependence of
$\rho_{xy}/\rho_{H}^{D}-1$ in the diffusive regime replaced by a
hyperbolic decrease $1/T$ at higher temperatures.

\begin{figure}
\includegraphics*[angle=-90,width=\columnwidth]{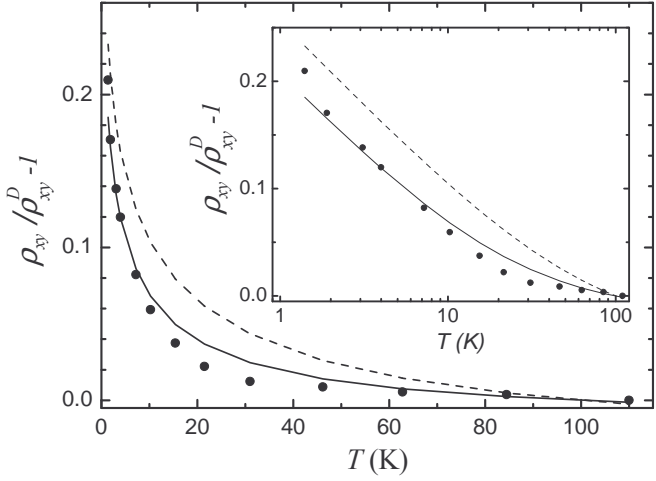}%
\caption{\label{Figure5} Temperature dependence of the Hall
coefficient (dots) compared to Eq.~(\ref{Eq3}) (dash line) and to
Eq.~(\ref{Eq3}) with $\lambda=0.065$ (solid line). The same data
are plotted in a logarithmic scale.}
\end{figure}

Figure~\ref{Figure5} shows how this prediction works in our case.
A simple calculation (carried out without any attempt at fitting
the experiment) results in the dashed curve ($F_0^\sigma=-0.17$).
This prediction is compared with the experimental correction
(black dots). At each temperatures the Hall coefficient was
obtained by linearly fitting the experimental curves shown in
Fig.~\ref{Figure1}b. The corresponding range of magnetic field
satisfies $\omega_c\tau<0.6-0.8,$ which allowed us to neglect
the finite-$B$ corrections to Eqs.~(\ref{Eq3})-(\ref{Eq4t}) in our analysis.
As shown in Ref.~\onlinecite{Gornyi}, such corrections are small even at
$\omega_c\tau \sim 1$
because of small numerical factors,  so that one can safely use
the results of Ref.~\onlinecite{Zala1} obtained in the limit $B\to 0$
in a rather wide range of $B$.

On the whole, there is a qualitative
agreement between theory and experiment but the quantitative
agreement is lacking. Using $F_0^\sigma$ as a fitting parameter
does not improve the agreement. Nevertheless we have found, that
if the coefficient $\lambda={11\pi\over 192}\simeq 0.18$ is
replaced by $\lambda={4\pi\over 192}\simeq 0.065$, then the
theoretical curve (the solid line) fits the experimental
dependence quite well.
\begin{figure}
\includegraphics*[angle=-90,width=\columnwidth]{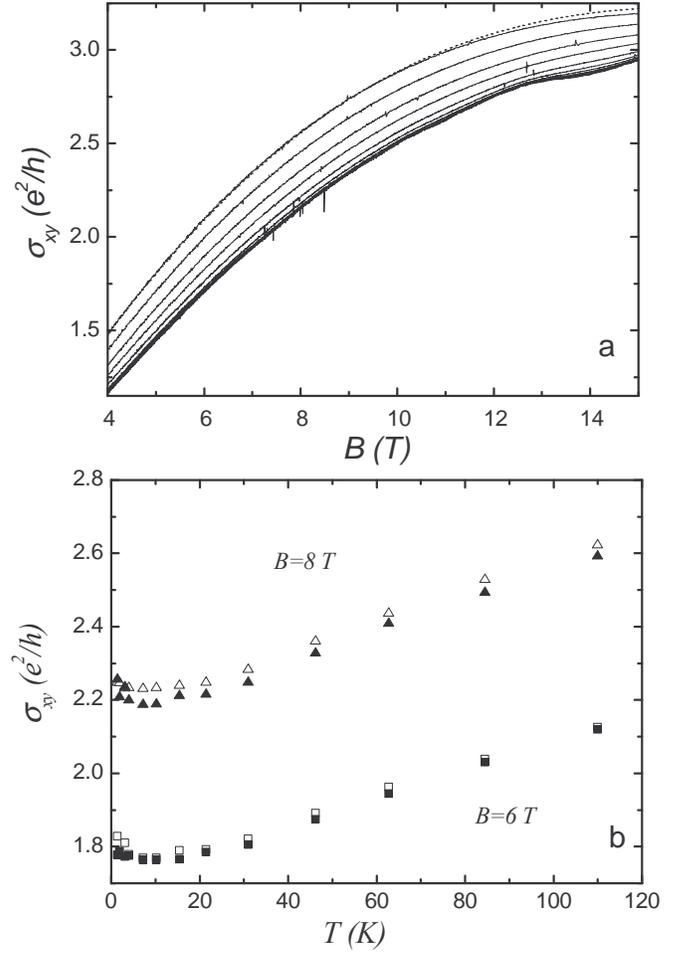}%
\caption{\label{Figure6} a) Transverse conductivity as function of
magnetic field (shown in the range of $B$ relevant to the
interaction-induced corrections) for the temperatures listed in the caption of
Fig.~\ref{Figure1}. The dotted line corresponds to
Eq.~(\ref{Eq5}) taken at $T=110$\,K;
b) Transverse conductivity as function of $T$ for
two different values of magnetic fields (black symbols). It is
compared to the value calculated using Eq.~(\ref{Eq5}) (Open
symbols).}
\end{figure}

This result might be related to an anisotropy of electron
scattering in the sample which reduces the electron return
probability and so weakens the correction at low fields
($\omega_c\tau\ll1$). The reduction of the pre-factor $\lambda$
could just be the way in which this anisotropy reveals itself
since the correction is proportional to $\lambda$ in the ballistic
limit. It is worth noting, that in the ballistic regime the
correction to the Hall coefficient is more sensitive to the
anisotropy of impurity scattering than the leading correction to
the longitudinal conductivity. This is because the relevant
processes giving rise to $\delta\rho_{xy}$ involve at least three
impurity scattering events, while those leading to the
linear-in-$T$ correction to $\sigma_{xx}$ involve a single
backscattering. Clearly, each large-angle scattering event yields
a reduction factor even for the weak anisotropy of scattering.\\
Finally, in Fig.~\ref{Figure6} we show the experimental data
points for the transverse conductivity tensor as a function of
magnetic field (Fig.~\ref{Figure6}a) and as function of
temperature for two different values of magnetic field
(Fig.~\ref{Figure6}b). The conductivity is observed to be
temperature independent at low temperatures and vary linearly with
temperature at high temperatures. While conform to the theoretical
prediction in the diffusive regime
($\Delta\sigma^\mathrm{H}_{ee}=0$ and $\Delta\rho_{xy}^{\mathrm{WL}}=0$,
according to Ref.~\onlinecite{Altshuler}), the behavior at high
temperatures is less obvious. However, this behavior follows from
Eq.~(\ref{Eq1-xy}) which takes into account the ballistic
renormalization of the scattering time. The measured values of the
Hall conductivity are indeed well compared to values of
$\sigma_{xy}(T)$ calculated using the simple Drude-like formula:
\begin{equation}
\label{Eq5}
\sigma_{xy}=\frac{e^2 n}{m^*} \frac{\omega_c\tau^2(T)}{1+\omega^2_c\tau^2(T)}.
\end{equation}
\noindent In this formula we neglected terms
$\omega_c\tau(T)\Delta\sigma^\mathrm{H}_{ee}(T)$ and
$\Delta\sigma_{xy}^{\mathrm{WL}}(T,B)$ from Eq.~(\ref{Eq1-xy}) and used
Eq.~(\ref{sigmaD}) for $\sigma_D(T)$. To evaluate $\sigma_{xy}$ we
used the values of the scattering time shown in
Fig.~\ref{Figure3}. Again the data are well described by the model
which includes no fitting parameter. Note that we also calculated
the expected field dependence at $T=110$\,K [see dotted curve in
(Fig.~\ref{Figure6}a)] which also reproduced well the experimental
data. A more detailed analysis of the Hall conductivity within the
general method outlined in Sec.~\ref{background} [taking into
account all the terms in Eq.~(\ref{Eq1-xy})] will be presented
elsewhere~\cite{unpub}.

\section{Longitudinal magnetoresistance}
\label{MR}

Let us return to the analysis of the longitudinal resistivity
$\rho_{xx}(B)$ shown in Fig.~\ref{Figure1}.
This analysis is aimed to obtain a consistent description
including all the transport coefficients,
$\rho_{xx}(B)$, $\sigma_{xx}(B)$,  $\rho_{xy}(B)$, and $\sigma_{xy}(B)$.
Furthermore, the behavior of $\rho_{xx}(B)$ in the ballistic regime
is determined by more subtle effects as compared to the behavior
of the conductivity tensor.
It turns out that in the longitudinal MR, unlike in the conductivity
components,
the leading $B$-independent e-e correction to $\tau$ cancels out.
In fact, the $T$ dependence of  $\rho_{xx}(B)$ reflects
the weak influence of magnetic field on the collision integral
in the quantum kinetic equation of Ref.~\onlinecite{Zala}.

\begin{figure}
\includegraphics*[angle=-90,width=\columnwidth]{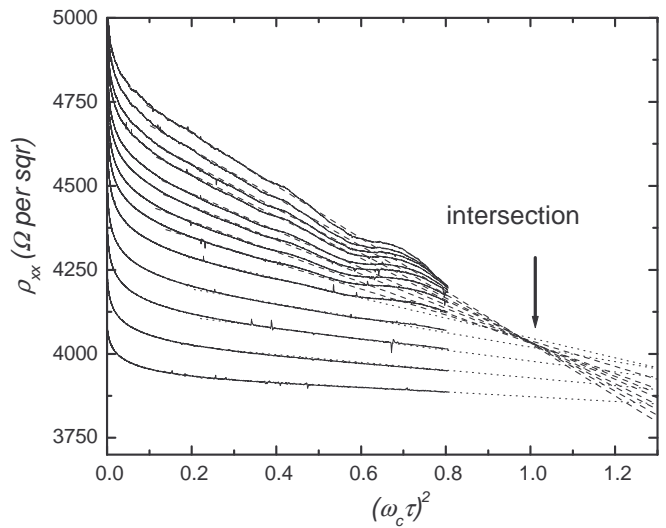}%
\caption{\label{Figure7} $\rho_{xx}$ plotted as a function of
$(\omega_c\tau)^2$ for the temperatures listed in the caption of
Fig.\ref{Figure1}. The dashed lines are the extrapolation of the
linear behavior of the curves corresponding to the diffusive regime.
They cross each other at $\omega_c\tau=1$. The dotted lines
represent the extrapolation of the curves in the ballistic regime.}
\end{figure}

As discussed in Sec. \ref{method},
the low-$B$ part of the curves is dominated by the WL-induced
MR, while the MR for $B>B_{tr}$ is  governed by the interaction effect.
In the diffusive regime $k_B T\tau/\hbar\ll 1$,
the interaction--induced resistivity correction,
\begin{eqnarray}
{\delta\rho_{xx}(B)\over \rho^D} &=& {{1-(\omega_c\tau)^2}\over 2 \pi E_F \tau}
\left(1+3\left[1-\frac{\ln(1+F_{0}^{\sigma})}{F_{0}^{\sigma}}\right]\right)\nonumber \\
&\times&
 \ln \left(\hbar\over k_B T\tau\right)
\label{MRAA}
\end{eqnarray}
($\rho^D$ is the classical Drude resistivity),
gives rise to a parabolic MR
$\Delta\rho_{xx}=\delta\rho_{xx}(B)-\delta\rho_{xx}(B=0)$
in arbitrary magnetic field~\cite{SenGir,girvin82}.

In the ballistic regime, as shown in Ref.~\onlinecite{Gornyi}, the
interaction--induced MR remains quadratic in magnetic field, while
the $T$ behavior of the proportionality coefficient depends on the
type of disorder. In the present case of short-ranged impurities,
Ref.~\onlinecite{Gornyi} predicts the following ballistic ($k_B
T\tau/\hbar\gg 1$) asymptotic behavior of the MR for
$\hbar\omega_c\ll 2\pi^2 k_B T$:
\begin{equation}
{\Delta\rho_{xx} \over \rho^D}= -(\omega_c\tau)^2
\frac{1+3g(F_{0}^{\sigma})}{2 \pi E_F \tau}  \frac{17 \pi \hbar}{192  k_B T\tau},
\label{rho-xx-ball}
\end{equation}
where the function $g(F_{0}^{\sigma})$ describes the contribution
of the triplet channel. It is worth stressing
that in high-density systems with $r_s\ll 1$
(i.e. for $|F_{0}^{\sigma}|\ll 1$), the parabolic MR is dominated
by the contribution of the singlet channel and hence is negative.

Equations (\ref{MRAA}) and (\ref{rho-xx-ball}) can be obtained by
inverting the conductivity tensor given by
Eqs.~(\ref{sigma_xx_no_WL}) and (\ref{Eq5}). One can see that the
classical part of the conductivity tensor [i.e. Eq.~(\ref{Eq5})
and the first term in Eqs.~(\ref{Eq1}) and (\ref{sigma_xx_no_WL})]
does not yield $B$ dependence of the resistivity, even when the
interaction effects are taken into account through the $T$
dependence of $\tau(T)$. Indeed, neglecting the term
$\Delta\sigma^\mathrm{Diff}_{ee}(T)$ one gets
$\rho_{xx}(T)=m^*/e^2 n \tau(T)$ which is independent of $B$. Thus
the MR is solely generated by the term
$\Delta\sigma^\mathrm{Diff}_{ee}$. We remind the reader that in
the ballistic regime $\Delta\sigma^\mathrm{Diff}_{ee}$ appears to
be dominated by the effect of magnetic field on the collision
integral, see Sec.~\ref{background}. Thus, we conclude that the
main source of the MR for $T>20-30$\,K is the weak $B$ dependence of
the transport scattering time.
\begin{figure}
\includegraphics*[angle=-90,width=\columnwidth]{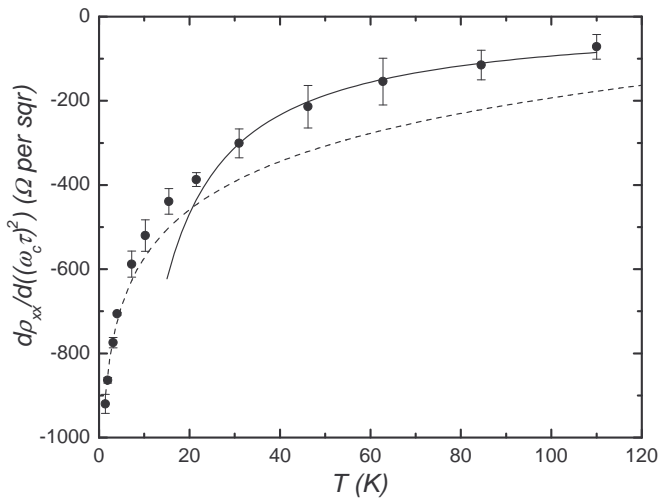}%
\caption{\label{Figure8} Slope of the curves shown in
Fig.~\ref{Figure7} compared to the theoretical  predictions. The
dashed line corresponds to the diffusive regime (Eq.~(\ref{MRAA}))
and the solid line to the ballistic limit
(Eq.~(\ref{rho-xx-ball})).}
\end{figure}

Let us now compare our experimental data with the above
theoretical predictions. It is worth mentioning that the
comparison is again parameter free. Figure~\ref{Figure7} presents
the longitudinal resistivity as a function of $(\omega_c\tau)^2$.
It shows that the MR is indeed parabolic and negative. The slope
of the curves $\rho_{xx}$ vs $(\omega_c\tau)^2$ was obtained in
the relevant ranges $(\omega_c\tau)^2 = 0.1-0.4$ for the curves
corresponding to $T<20$\,K
and for $(\omega_c\tau)^2>0.2$ for $T>20$\,K. This has allowed us to
reduce the influence of WL in the high $T$ data and the SdH
oscillations in the low $T$ data. The slope of these lines is
presented in figure \ref{Figure8}. The error in the determination
of the slope due to the choice of the evaluation range was
estimated from the deviations obtained using the interval
$0.2-0.35$
of $(\omega_c\tau)^2$ for the linear fit. In Fig.~\ref{Figure7} we
have extrapolated the MR lines to the region of higher magnetic
fields $\omega_c\tau\sim 1$. From this plot we again clearly see
the crossover between the diffusive and the ballistic regimes
which occurs at $T\sim 20$\,K. Indeed the lines for $T<20$\,K
intersect each other at a single point close to
$(\omega_c\tau)^2=1$, as predicted by the diffusive expression
(\ref{MRAA}). As follows from Eq.~(\ref{MRAA}), at the
intersection point the quantum correction to the longitudinal
resistivity is zero, so that the value of $\rho_{xx}$ at this
point corresponds to the classical Drude value of the resistivity:
\begin{equation}
\rho_{xx}(\omega_c\tau=1)=\rho^D=1/\sigma_0^D.
\label{Drude-rho}
\end{equation}
This value of $\sigma_0^D$ was used in Sec.~\ref{method} to find
the magnitude of the interaction-induced conductivity correction
at $B=0$. For higher temperatures ($T>20$\,K), the MR lines in
Fig.~7 no longer intersect each other at a single point. At this
point the system enters the crossover region, where the
$T$-dependence of $\sigma_D(T)$ starts to become important.

The proportionality coefficient of $\rho_{xx}$ vs
$(\omega_c\tau)^2$ dependence is compared in Fig.~\ref{Figure8}
to the theoretical asymptotics given by Eqs.~(\ref{MRAA})
and (\ref{rho-xx-ball}). In (\ref{MRAA}) we used the ``diffusive''
value $F_0^\sigma=-0.17$ given by Eq.~(\ref{F-diff}). In
Eq.~(\ref{rho-xx-ball}) we used
$g(F_{0}^{\sigma})=F_{0}^{\sigma}/(1+F_{0}^{\sigma})$ with
$F_{0}^{\sigma}=-0.1$ given by Eq.~(\ref{F-ball}). This is
consistent with the above observation that the ``ballistic'' MR is
mostly due to the $B$-dependent corrections to the collision
integral. An almost perfect quantitative agreement between the
predictions of Refs.~\onlinecite{SenGir}, \onlinecite{girvin82}
and \onlinecite{Gornyi} and the experimental data is found for
both diffusive and ballistic temperature regimes.

\section{Conclusion}

In conclusion, we have presented a study aiming at observing the
crossover from the diffusive to the ballistic regime in the weak
interaction limit. We find strong evidences of such crossover in
the obtained measurements. We realized a parameter free comparison
of our experimental data for the longitudinal conductivity and
Hall coefficient to the recent ZNA theory as well as the longitudinal resistivity
to the theory of Ref.~\onlinecite{Gornyi}. We find these
theories to be in a good qualitative agreement with our experimental
results.

\begin{acknowledgments}
We are very grateful to A. Dmitriev, A. Germanenko, G. Minkov, A.
Mirlin, and B. Narozhny for useful discussions. This work was
supported by PICS-RFBR (Grant No 1577.), RFBR (Grants No
02-02-16516, 05-02-17800, and 05-02-17802), NATO, INTAS (Grant No
01-0014), DFG-Schwerpunktprogramm ``Quanten-Hall-Systeme",
programs ``Physics and Technology of Nanostructures" of the
Russian ministry of Industry and Science, ``Low dimensional
quantum structures" of RAS, and ``Russian Scientific School"
(Grant No 2192.2003.2).

\end{acknowledgments}



\begin{thebibliography}{99}
\bibitem{foot-WL} A subtle point is that the residual $T$-independent ``ballistic''
WL contribution decreases rather slowly ($\propto B^{-1/2}$)
at $B>B_{tr}$. As discussed in Refs.~\onlinecite{Minkov,Minkov2}, in relatively
low-mobility structures this WL tail may be still important for the analysis
of magnetotransport data, when the fitting is performed in the range $B<10 B_{tr}$.
However, this $T$-independent WL term drops out in Eq.~(\ref{difference}) and
therefore the interaction correction $\Delta\sigma_{xx}^{ee}(T,B=0)$ obtained
from Eq.~(\ref{difference}) is almost unaffected by the WL tail.
\end{thebibliography}
\end{document}